\documentclass[apj]{emulateapj}

\def\gtorder{\mathrel{\raise.3ex\hbox{$>$}\mkern-14mu
             \lower0.6ex\hbox{$\sim$}}}
\def\ltorder{\mathrel{\raise.3ex\hbox{$<$}\mkern-14mu
             \lower0.6ex\hbox{$\sim$}}}

\slugcomment{Draft of \today}

\shorttitle{Supernova Collisionless Shocks}

\shortauthors{Ofek et al.}

\begin{document}

\title{X-ray emission from supernovae in dense circumstellar matter environments: A search for collisionless shocks}
\author{E.~O.~Ofek\altaffilmark{1},
D.~Fox\altaffilmark{2}
S.~B.~Cenko\altaffilmark{3},
M.~Sullivan\altaffilmark{4},
O.~Gnat\altaffilmark{5},
D.~A.~Frail\altaffilmark{6},
A.~Horesh\altaffilmark{7},
A.~Corsi\altaffilmark{8},
R.~M.~Quimby\altaffilmark{9},
N.~Gehrels\altaffilmark{10},
S.~R.~Kulkarni\altaffilmark{7},
A.~Gal-Yam\altaffilmark{1},
P.~E.~Nugent\altaffilmark{11},
O.~Yaron\altaffilmark{1},
A.~V.~Filippenko\altaffilmark{3},
M.~M.~Kasliwal\altaffilmark{12},
L.~Bildsten\altaffilmark{13}$^{,}$\altaffilmark{14},
J.~S.~Bloom\altaffilmark{3},
D.~Poznanski\altaffilmark{15},
I.~Arcavi\altaffilmark{1},
R.~R.~Laher\altaffilmark{16},
D.~Levitan\altaffilmark{7},
B.~Sesar\altaffilmark{7},
J.~Surace\altaffilmark{16}
}

\altaffiltext{1}{Benoziyo Center for Astrophysics, Weizmann Institute
  of Science, 76100 Rehovot, Israel.}
\altaffiltext{2}{Department of Astronomy \& Astrophysics, Pennsylvania State University, University Park, PA 16802}
\altaffiltext{3}{Department of Astronomy, University of California,
  Berkeley, Berkeley, CA 94720-3411.}
\altaffiltext{4}{Department of Physics, University of Oxford, Denys
  Wilkinson Building, Keble Road, Oxford OX1 3RH, UK.}
\altaffiltext{5}{Racah Institute of Physics, The Hebrew University of Jerusalem 91904, Israel}
\altaffiltext{6}{National Radio Astronomy Observatory, P.O. Box O, Socorro, NM 87801}
\altaffiltext{7}{Division of Physics, Mathematics and Astronomy,
  California Institute of Technology, Pasadena, CA 91125.}
\altaffiltext{8}{LIGO laboratory, Division of Physics, California Institute of Technology, MS 100-36, Pasadena, CA 91125}
\altaffiltext{9}{Kavli IPMU, University of Tokyo, 5-1-5 Kashiwanoha, Kashiwa-shi, Chiba, 277-8583, Japan}
\altaffiltext{10}{NASA-Goddard Space Flight Center, Greenbelt, Maryland 20771}
\altaffiltext{11}{Lawrence Berkeley National Laboratory, 1 Cyclotron
  Road, Berkeley, CA 94720.}
\altaffiltext{12}{Observatories of the Carnegie Institution for Science, 813 Santa Barbara St, Pasadena CA 91101 USA}
\altaffiltext{13}{Kavli Institute for Theoretical Physics, Kohn Hall,
  University of California, Santa Barbara, CA 93106.}
\altaffiltext{14}{Department of Physics, Broida Hall, University of
  California, Santa Barbara, CA 93106.}
\altaffiltext{15}{School of Physics and Astronomy, Tel-Aviv University, Israel}
\altaffiltext{16}{Spitzer Science Center, California Institute of Technology,  M/S 314-6, Pasadena, CA 91125}

\begin{abstract}


The optical light curve of some supernovae (SNe)
may be powered by the outward diffusion of the energy
deposited by the explosion shock
(so-called shock breakout)
in optically thick ($\tau\gtorder 30$)
circumstellar matter (CSM).
Recently, it was shown that the radiation-mediated
and -dominated shock in an optically thick wind
must transform into a collisionless shock
and can produce hard X-rays.
The X-rays are expected to peak
at late times, relative to maximum visible light.
Here we report on a search,
using {\it Swift}-XRT and {\it Chandra},
for X-ray emission from
28 SNe that belong to classes
whose progenitors are suspected
to be embedded in dense CSM.
Our sample includes 
19 type-IIn SNe, one type-Ibn SN and eight
hydrogen-poor super-luminous SNe
(SLSN-I; SN\,2005ap like).
Two SNe (SN\,2006jc and SN\,2010jl)
have X-ray properties that are
roughly consistent with the expectation for
X-rays from a collisionless shock in optically thick CSM.
Therefore, we suggest that their optical light curves
are powered by shock breakout in CSM.
We show that two other events (SN\,2010al and SN\,2011ht)
were too X-ray bright during the SN maximum optical light
to be explained by the shock breakout model.
We conclude that the light curves of some, but not all,
type-IIn/Ibn SNe are powered by shock breakout in CSM.
For the rest of the SNe in our sample, including all the SLSN-I events,
our X-ray limits are not deep enough and were typically
obtained at too early times (i.e., near the SN maximum light)
to conclude about their nature.
Late time X-ray observations
are required in order to further test
if these SNe are indeed embedded in dense CSM.
We review the conditions required for
a shock breakout in a wind profile.
We argue that the time scale, relative to maximum light,
for the SN to peak in X-rays is a probe of the
column density and the density profile above the
shock region.
The optical light curves of SNe,
for which the X-ray emission peaks at late times, are likely powered
by the diffusion of shock energy from
a dense CSM.
We note that if the CSM density profile falls faster
than a constant-rate wind density profile,
then X-rays may escape at earlier times than estimated
for the wind profile case.
Furthermore, if the CSM have a region
in which the density profile is very steep,
relative to a steady wind density profile,
or the CSM is neutral,
then the radio free-free absorption 
may be low enough, and radio emission may be detected.

\end{abstract}


\section{Introduction}
\label{sec:Introduction}

Circumstellar Matter (CSM) around supernova (SN)
progenitors may play an important role in the emission
and propagation of energy from SN explosions.
The interaction of the SN radiation with optically thin CSM shells
may generate emission lines, with widths that are representative
of the shell velocity (i.e., type-IIn SNe; Schlegel 1990; Kiewe et al.\ 2012).
The interaction of SN ejecta with the CSM can
power the light curves of SNe by
transformation of the SN kinetic energy into photons.
In cases where considerable amounts of optically thin
(and ionized) material is present around the exploding star,
synchrotron and free-free radiation
can emerge, and inverse Compton scattering can generate
X-ray photons
(e.g., Chevalier \& Fransson 1994; Horesh et al.\ 2012;
Krauss et al.\ 2012).

For the type-IIn SN PTF\,09uj, Ofek et al.\ (2010)
suggested that a shock breakout can
take place in optically-thick wind
(see also Grassberg, Imshennik, \& Nadyozhin 1971;
Falk \& Arnett 1977;
Chevalier \& Irwin 2011;
Balberg \& Loeb 2011).
This will happen if the Thomson optical depth
within the wind profile
is $\gtorder c/v_{{\rm sh}}$, where $c$ is the speed of light,
and $v_{{\rm sh}}$ is the shock speed.
Ofek et al.\ (2010)
showed that shock breakout in wind environments produces optical
displays that are brighter and have longer time scales than those
from surfaces of red supergiants
(e.g., Colgate 1974;
Matzner \& McKee 1999;
Nakar \& Sari 2010;
Rabinak \& Waxman 2011;
Couch et al.\ 2011).
Chevalier \& Irwin (2011) extended this picture.
Specifically, they discussed CSM with a wind profile
in which the wind 
has a cutoff at a distance $R_{{\rm w}}$.
If the optical depth at $R_{{\rm w}}$
is $\ltorder c/v_{{\rm s}}$ then
the light curve of the supernova will have a slow decay
(e.g., SN\,2006gy; Ofek et al.\ 2007; Smith et al.\ 2007).
If the optical depth at $R_{{\rm w}}$
is $\gtorder c/v_{{\rm s}}$, then
it will have a faster decay
(e.g., SN\,2010gx; Pastorello et al.\ 2010a; Quimby et al.\ 2011a).
Moriya \& Tominaga (2012) investigated
shock breakouts in general wind density profiles
of the form $\rho\propto r^{-w}$.
They suggested that,
depending on the power-law index $w$,
shock breakouts in wind environments
can produce bright SNe without narrow emission
lines (e.g., SN\,2008es; Gezari et al.\ 2009; Miller et al.\ 2009).

Recently Katz et al.\ (2011) and Murase et al. (2011) showed that
if the progenitor is surrounded by optically thick
CSM then a collisionless shock is necessarily formed
during the shock breakout.
Moreover, they argued that the energy emitted
from the collisionless shock
in the form of high-energy photons and particles
is comparable to the shock breakout energy.
Furthermore, this process may generate high-energy ($\gtorder 1$\,TeV)
neutrinos.
Although Katz et al.\ (2011) predicted that the photons
are generated with energy typically above 60\,keV, it
is reasonable to assume that some photons will be emitted
with lower energy.
Chevalier \& Irwin (2012) showed that Comptonization
and inverse Compton
of the high-energy photons is likely to play
an important role, and that the high-energy photons
will be absorbed.

Svirski, Nakar \& Sari (2012) discuss the X-ray emission
from collisionless shocks.
They show that at early times the X-rays will
be processed into the optical regime by
the Compton process.
Therefore, at early times, the optical emission
will be about $10^{4}$ times stronger than the high-energy emission.
With time, the X-ray emission will become stronger,
while the optical emission will decay.
They conclude that for a CSM with a steady
wind profile ($w=2$), X-ray emission may peak
only at late times, roughly 10--50 times
the shock breakout time scale.
The shock breakout time scale, $t_{{\rm br}}$,
is roughly given by the diffusion time scale
at the time of shock breakout.
This time scale is also equivalent to the
radius at which the shock breaks out ($r_{{\rm br}}$)
divided by the shock velocity ($v_{{\rm s}}$; Weaver 1976).
If the main source of optical photons is
due to diffusion of the shock breakout energy,
the SN optical light rise time, $t_{{\rm rise}}$,
will be equivalent to the shock breakout time scale.
Therefore, X-ray flux measurements and spectra of
SNe embedded in dense CSM starting
from the explosion until months or years after maximum
light are able to measure the properties of the CSM
around the SN progenitors and the progenitor mass-loss history.
This unique probe into the final stages of massive star
evolution has been only partially exploited, at best.

Herein, we analyze the X-ray data for 28 SNe with light curves
that may be powered by a shock breakout from dense CSM,
and for which {\em Swift}-XRT
(Gehrels et al.\ 2004) observations exist.
We use this sample to search for X-ray signatures of
collisionless shocks -- emission
at late times (months to years after peak optical luminosity).
We suggest that these signals were observed in
several cases, most notably in SN\,2006jc
(Immler et al.\ 2008) and SN\,2010jl (Chandra et al.\ 2012a).
Finally, we review the conditions for a shock breakout
in CSM with a wind profile
and discuss, the importance of bound-free absorption
and the possibility to detect radio emission
from such SNe.

The structure of this paper is as follows.
in \S\ref{sec:Sample} we present the SN sample,
while \S\ref{sec:Obs} presents the X-ray observations.
We review and discuss the model in \S\ref{sec:Disc},
and discuss the observations in context of the model in \S\ref{sec:obsmoel}.
Finally, we conclude in \S\ref{Conc}.

\section{Sample}
\label{sec:Sample}

Our sample of SNe is based on SNe found by amateur astronomers,
and several surveys, including the
Lick Observatory Supernova Search (LOSS; Li et al.\ 2000),
the Catalina Real-Time Transient Survey (CRTS; Drake et al.\ 2009a),
Pan-STARRS1 (PS1; Kaiser et al.\ 2002),
and the Palomar Transient Factory\footnote{http://www.astro.caltech.edu/ptf/}
(PTF; Law et al.\ 2009; Rau et al.\ 2009).
Two SNe, PTF\,09drs and PTF\,10tel, are reported here
for the first time.
We note that many of the nearby or luminous SNe found by
PTF are also observed by {\it Swift}.

We selected a sample of SNe in which the main source of energy
may be explained by diffusion of the explosion shock
energy through optically thick CSM 
around the progenitor.
First, we include type-IIn SNe within 200\,Mpc.
SNe that belong to this class show narrow hydrogen emission lines.
This is an indication of the presence of optically thin
material somewhere around the progenitor.
However, it is unlikely that {\it all} SNe
showing narrow hydrogen emission lines in their spectra
are powered mainly
by the diffusion of shock energy in an optically thick environment.
One reason is that some type-IIn SNe show X-ray emission
near maximum optical light, which is not expected
when optically thick CSM is present (see \S\ref{sec:Disc}).
Furthermore, Moriya \& Tominaga (2012) suggested
that not all SNe powered by interaction
of the ejecta with slow-moving material
will necessarily have narrow emission lines in their spectrum.
We note that some of the type-IIn SNe in our sample
are peculiar (e.g., SN\,2010jp/PTF\,10aaxi; Smith et al.\ 2012).

Another relevant, but rare, class of objects are type-Ibn SNe.
This class is defined by the lack of hydrogen lines
and presence of narrow Helium emission lines.
The only SN of this type in our sample is SN\,2006jc
(Nakano et al.\ 2006; Foley et al.\ 2007; Pastorello et al.\ 2008).

The third class of SNe we investigate here
is the small group of hydrogen-poor super-luminous SN
(SLSN-I; see review in Gal-Yam et al.\ 2012).
Quimby et al.\ (2011a) used the spectra of several such
events found by PTF, at intermediate redshift ($z\sim0.5$),
to show that these events, as well as
SCP\,06F6 (Barbary et al.\ 2009), and
SN\,2005ap (Quimby et al.\ 2007) are spectroscopically similar.
This group of SNe continues to grow with
new discoveries
(e.g., Chomiuk et al.\ 2011; Leloudas et al.\ 2012),
and their hosts were studied in Neill et al.\ (2011).
Although the nature of these events is not understood
(e.g., Kasen \& Bildsten 2010),
Quimby et al.\ (2011a) suggested that they may be powered by a
pulsational pair-instability SN (Rakavy, Shaviv, \& Zinamon 1967;
Woosley, Blinnikov, \& Heger 2007).
According to this tentative model,
the SN ejecta interacts with a dense shell of material,
enriched with intermediate mass elements, that was ejected
by the progenitor during previous explosions
(see also Ginzburg \& Balberg 2012).
This model is tentatively supported by observations
of SN\,2006oz (Leloudas et al.\ 2012) that may show
a dip in the light curve followed by re-brightening.
Moriya \& Maeda (2012) interpret the dip as an increase
in the opacity due to ionization
of the massive shell/CSM as it interact with the ejecta.

Our sample, presented in Table~\ref{tab:Samp},
consists of eight SLSN-I objects,
19 type-IIn SNe and a single type-Ibn SN.
We note that the spectra of SNe with PTF names,
as well as some other SNe,
are available online from the
WISeREP\footnote{Weizmann\ Interactive Supernova (data) REPository; http://www.weizmann.ac.il/astrophysics/wiserep/} website
(Yaron \& Gal-Yam 2012).

\begin{deluxetable*}{llrrlllllll}
\tablecolumns{11}
\tablewidth{0pt}
\tablecaption{SN sample}
\tablehead{
\colhead{Name}          &
\colhead{Type}          &
\colhead{$\alpha_{{\rm J2000}}$}      &
\colhead{$\delta_{{\rm J2000}}$}     &
\colhead{$t_{{\rm rise}}$} &
\colhead{$M_{{\rm R}}$} &
\colhead{z}             &
\colhead{$t_{{\rm peak}}$} &
\colhead{N$_{{\rm H}}$}   &
\colhead{$L_{{\rm X}}$}   &
\colhead{$L_{{\rm X}}/L_{{\rm opt}}$} \\
\colhead{}              &
\colhead{}              &
\colhead{deg}           &
\colhead{deg}           &
\colhead{day}           &
\colhead{mag}           &
\colhead{}              &
\colhead{MJD}           &
\colhead{$10^{20}$\,cm$^{-2}$} &
\colhead{erg\,s$^{-1}$}  &
\colhead{}
}
\startdata
PTF\,09atu             & SLSN-I & 247.602288 & $+$23.640289 & 30:        &$-$22.5  & 0.501  & 55060  & 4.05 & $<1.9\times10^{44}$        & 0.7\\
PTF\,09cnd             & SLSN-I & 243.037252 & $+$51.487820 & 50         &$-$22.8  & 0.258  & 55080  & 1.67 & $<8.0\times10^{42}$        & 0.02\\
PTF\,09cwl/SN\,2009jh  & SLSN-I & 222.291998 & $+$29.419833 & 50         &$-$22.5: & 0.349  & 55060  & 1.51 & $<1.1\times10^{44}$        & 0.4\\
SN\,2010gx/PTF\,10cwr  & SLSN-I & 171.444479 & $-$8.828099  & 20         &$-$21.7: & 0.231  & 55280  & 3.78 & $<9.9\times10^{42}$        & 0.07\\
PTF\,10hgi             & SLSN-I & 249.446009 & $+$6.208978  & 50         &$-$20.3  & 0.096  & 55370  & 6.06 & $<5.1\times10^{42}$        & 0.1\\
PTF\,11dij             & SLSN-I & 207.740690 & $+$26.278562 & 40:        &$-$21.1: & 0.143  & 55690  & 1.21 & $<4.6\times10^{42}$        & 0.06\\
PTF\,11rks             & SLSN-I &  24.939618 & $+$29.924170 & 20         &$-$21.0  & 0.20   & 55945  & 5.27 & $<5.4\times10^{42}$        & 0.07\\
PS1-12fo               & SLSN-I & 146.553792 & $+$19.841306 & $>14$      &$-$21.0: & 0.175  & 55956  & 2.79 & $<1.8\times10^{43}$        & 0.2\\
SN\,2006jc             & Ibn    & 139.366667 & $+$41.908889 & $<15$      &$-$17.8  & 0.006  & 54020  & 1.00 & $\approx1.5\times10^{41}$  & 0.04\\
PTF\,09drs             & IIn    & 226.625665 & $+$60.594271 & 40:        &$-$17.8: & 0.045  & 55210  & 1.72 & $<4.4\times10^{42}$        & 1.2\\
SN\,2010jl/PTF\,10aaxf & IIn    & 145.722208 & $+$9.494944  &$19\pm4$    &$-$20.6  & 0.011  & 55500  & 3.05 & $\approx1.8\times10^{41}$  & 0.004\\
SN\,2010jp/PTF\,10aaxi & IIn    &  94.127702 & $-$21.410012 &$<19$       &$-$14.6: & 0.01   & 55520  & 11.0 & $<1.2\times10^{40}$        & 0.06\\
SN\,2010jj/PTF\,10aazn & IIn    & 31.717743  & $+$44.571558 &$34\pm19$   &$-$18.0: & 0.016  & 55530  & 9.38 & $<1.2\times10^{41}$        & 0.03\\
SN\,2010bq/PTF\,10fjh  & IIn    & 251.730659 & $+$34.159641 & $15..45$   &$-$18.5  & 0.032  & 55310  & 1.79 & $<1.2\times10^{42}$        & 0.2\\
PTF\,11iqb             & IIn    &   8.520150 & $-$9.704979  & 10:        &$-$18.4  & 0.013  & 55780  & 2.79 & $\approx7.9\times10^{40}$  & 0.01\\
SN\,2007bb             & IIn    & 105.281083 & $+$51.265917 &$<15$       &$-$17.6: & 0.021  & 54192  & 7.04 & $<2.8\times10^{41}$        & 0.09\\
SN\,2007pk             & IIn    &  22.946125 & $+$33.615028 &$<14$       &$-$17.3: & 0.017  & 54423  & 4.72 & $<1\times10^{41}$          & 0.04\\
SN\,2008cg             & IIn    & 238.563125 & $+$10.973611 &$45\pm15$   &$-$19.4: & 0.036  & 54583: & 3.65 & $<2.6\times10^{41}$        & 0.02\\
SN\,2009au             & IIn    & 194.941667 & $-$29.602083 &\nodata     &$-$16.5: & 0.009  & 54901: & 6.42 & $<3.5\times10^{40}$        & 0.03\\
SN\,2010al             & IIn    & 123.566292 & $+$18.438389 &$<35$       &$-$16.0: & 0.0075 & 55268: & 3.92 & $\approx2.2\times10^{41}$  & 0.3\\
SN\,2011ht             & IIn    & 152.044125 & $+$51.849167 &30          &$-$14.2: & 0.004  & 55880  & 0.78 & $\approx7.2\times10^{39}$  & 0.05\\
SN\,2011hw             & IIn    & 336.560583 & $+$34.216417 &\nodata     &$-$19.1: & 0.023  & 55883: & 10.2 & $<5.1\times10^{40}$        & 0.004\\
PTF\,10tel             & IIn    & 260.377817 & $+$48.129834 &17          &$-$18.5  & 0.035  & 55450  & 2.34 & $<7.2\times10^{41}$        & 0.1\\
SN\,2011iw             & IIn    & 353.700833 & $-$24.750444 &$<40$       &$-$18.1: & 0.023  & 55895: & 1.61 & $<1.0\times10^{41}$        & 0.02\\
SN\,2005db             & IIn    &  10.361625 & $+$25.497667 &$<18$       &$-$16.8: & 0.0153 & 53570: & 4.17 & $\approx5.3\times10^{40}$  & 0.04\\
SN\,2005av             & IIn    & 311.156583 & $-$68.752944 &$<19$       &$-$17.8: & 0.0104 & 53453: & 4.85 & $<8.1\times10^{39}$        & 0.02\\
SN\,2003lo             & IIn    &  54.271333 & $-$5.038139  &\nodata     &$-$15.8: & 0.0079 & 53005: & 4.87 & $<6.3\times10^{39}$        & 0.01\\
SN\,2002fj             & IIn    & 130.187917 & $-$4.127361  &$<90$       &$-$18.5  & 0.0145 & 52532  & 3.12 & $<4.9\times10^{40}$        & 0.007
\enddata
\tablecomments{The sample of SNe with X-ray observations.
Type refer to SN type,
$\alpha_{{\rm J2000}}$ and $\delta_{{\rm J2000}}$ are the J2000.0 right ascension
and declination, respectively.
$t_{{\rm rise}}$ is the approximate rise time of the SN optical light curve.
The rise time is deduced from various sources including
PTF and KAIT photometry and the literature listed in the references.
We note that $\pm$ sign indicates the range
rather than the uncertainty in $t_{{\rm rise}}$. The column sign indicates uncertain value.
$M_{{\rm R}}$ is the approximate absolute $R$-band magnitude at maximum light
(ignoring $k$-corrections).
$z$ refers to the object redshift. If the galaxy is nearby and has a direct distance measurements in the NASA Extragalactic Database (NED),
we replaced the observed redshift by the redshift corresponding
to the luminosity distance of the galaxy.
$t_{{\rm peak}}$ is the MJD of maximum light, and
$N_{{\rm H}}$ is the Galactic neutral hydrogen
column density for the source position
(Dickey \& Lockman 1990).
$L_{{\rm X}}$ is the X-ray luminosity or the 2-$\sigma$ upper
limit on the X-ray luminosity in the 0.2--10\,keV band.
Finally, $L_{{\rm X}}/L_{{\rm opt}}$ is the ratio between the X-ray measurments or limit ($L_{{\rm X}}$)
and the peak visible light luminosity.
\\
{\it References:}\\
PTF\,09atu: Quimby et al.\ (2011a). \\
PTF\,09cnd: Quimby et al.\ (2011a); Chandra et al.\ (2009); Chandra et al.\ (2010). \\
PTF\,09cwl: SN\,2009jh; Quimby et al.\ (2011a); Drake et al.\ (2009b). \\
SN\,2010gx: PTF\,10cwr; Mahabal et al.\ (2010); Quimby et al.\ (2010a); Pastorello et al.\ (2010b); Quimby et al.\ (2011). \\
PTF\,10hgi: Quimby et al.\ (2010b). \\
PTF\,11dij: Drake et al.\ (2011a); Drake et al.\ (2011b); Quimby et al.\ (2011b). \\
PTF\,11rks: Quimby et al.\ (2011c) \\
PS1-12fo:   Drake et al.\ (2012); Smartt et al.\ (2012); Maragutti et al.\ (2012). \\
PTF\,09drs: Reported here for the first time. \\
SN\,2010jl: PTF\,10aaxf; Newton et al.\ (2010); Stoll et al.\ (2011). \\
SN\,2010jp: PTF\,10aaxi; A peculiar type-IIn supernova; Maza et al.\ (2010); Challis et al.\ (2010b); Smith et al.\ (2012). \\
SN\,2010jj: PTF\,10aazn; Rich (2010b); Silverman et al.\ (2010b). \\
SN\,2010bq: PTF\,10fjh; Duszanowicz et al.\ (2010); Challis et al.\ (2010a). \\
PTF\,11iqb: Parrent et al.\ (2011); Horesh et al.\ (2011). \\
SN\,2007bb: Joubert \& Li(2007); Blondin et al.\ (2007). $t_{{\rm peak}}$ and $t_{{\rm rise}}$ are based on unpublished KAIT photometry.\\
SN\,2007pk: A peculiar type-IIn supernova (Parisky et al.\ 2007). $t_{{\rm peak}}$ and $t_{{\rm rise}}$ are based on unpublished KAIT photometry.\\
SN\,2008cg: Drake et al.\ (2008); Blondin et al.\ (2008); Filippenko et al.\ (2008; cbet 1420); Spectrum is similar to SN\,1997cy (Filippenko 2008). \\
SN\,2009au: Pignata et al.\ (2009); Stritzinger et al.\ (2009). \\
SN\,2010al: Spectrum is similar to SN\,1983K with \ion{He}{2}, \ion{N}{3}, and \ion{H}{1} emission lines; Rich et al.\ (2010a); Stritzinger et al.\ (2010); Silverman et al.\ (2010a). \\
SN\,2011ht: Boles et al.\ (2011); Prieto et al.\ (2011); Roming et al (2012).\\ 
SN\,2011hw: Dintinjana et al.\ (2011). \\ 
PTF\,10tel: Reported here for the first time. \\
SN\,2011iw: Mahabal et al.\ (2011). \\
SN\,2005db: Blanc et al.\ (2005); Monard (2005b); Kiewe et al.\ 2012). \\
SN\,2005av: Monard (2005a); Salvo et al.\ (2005). \\
SN\,2003lo: Puckett et al.\ (2004); Matheson et al.\ (2004). \\
SN\,2002fj: Monard \& Africa (2002); Hamuy (2002).
}
\label{tab:Samp}
\end{deluxetable*}

\section{Observations}
\label{sec:Obs}

For each {\em Swift}-XRT image of an SN,
we extracted the number of X-ray counts in the 0.2--10\,keV
band within an aperture of $7.2''$ (3\,pixels) radius
centered on the SN position.
We chose small aperture in order to minimize
any host galaxy contamination.
We note that this aperture contains $\approx37$\% of
the source flux (Moretti et al.\ 2004).
The background count rates were estimated in
annuli around each SN, with an inner (outer) radius of $50''$ ($100''$).
For each SN that has {\it Swift}-XRT observations,
we searched for {\it Chandra} observations.
The {\it Chandra} observations were analyzed in a similar
manner with an extraction aperture of $2''$
and background annuli with an inner (outer) radius of $15''$ ($40''$).
All the {\it Swift}-XRT X-ray measurements are listed in Table~\ref{tab:Obs}
(the full table is available in the electronic version).
In addition, in Table~\ref{tab:Obs},
for each object we give the count rate in up to four
super-epochs:
(i) all the observations taken prior to maximum light, or discovery date if time
of maximum light is not known;
(ii) all the observations taken between maximum light and 300 days after
maximum light;
(iii) all the observations taken more than 300 days after maximum light;
and (iv) all the observations at the position of the SN.

In each epoch, and super-epoch, we also provide the source false alarm
probability (FAP), which is the probability
that the X-ray counts are due to the X-ray background rather than
a source.
This probability is estimated as
1 minus the Poisson cumulative distribution
to get a source count rate smaller than the observed
count rate, assuming the expectancy value of the Poisson
distribution equal to the background
counts\footnote{Typically the background level is known very well.}
within an area equivalent to the aperture area.
We note that in some cases X-ray emission from the host galaxy
will tend to produce some seemingly significant, but actually "false alarm"
detections under these assumptions.
In cases in which FAP$\le0.01$,
we estimated also the 2-$\sigma$ upper limit on the count rate
(Gehrels 1986).
The count-rate measurements or upper limits are
converted to luminosity in the 0.2--10\,keV band
using the PIMMS\footnote{http://cxc.harvard.edu/toolkit/pimms.jsp} web tool and
assuming that:
the aperture in which we extracted the source
photometry contains 37\% of the source photons (Moretti et al.\ 2004);
Galactic neutral hydrogen column density in the position of the sources
as listed in Table~\ref{tab:Samp} (Dickey \& Lockman 1990);
a spectrum of $N_{{\rm ph}}(E)\propto E^{-0.2}$,
where $N(E)$ has units of photons\,cm$^{-2}$\,s$^{-1}$
(motivated in \S\ref{sec:Disc});
and the luminosity distance to the SNe
calculated using the redshifts listed
in Table~\ref{tab:Samp} 
and 
$H_{0}=70.4$\,km\,s$^{-1}$\,Mpc$^{-1}$; $\Omega_{m}=0.268$; $\Omega_{\Lambda}=0.716$; $\Omega_{K}=0.986$ (the 3rd year WMAP$+$SNLS cosmology;
Spergel et al.\ 2007).

\begin{deluxetable*}{lrrrllllll}
\tablecolumns{10}
\tablewidth{0pt}
\tablecaption{{\it Swift}-XRT X-ray measurements}
\tablehead{
\colhead{Name}          &
\multicolumn{2}{c}{$t-t_{{\rm peak}}$} &
\colhead{Exp.}          &
\colhead{CR}            &
\colhead{$\Delta$CR$_{-}$}       &
\colhead{$\Delta$CR$_{+}$}       &
\colhead{CR$_{{\rm UL}}$}  &
\colhead{FAP}            &
\colhead{$L_{{\rm X}}$}    \\
\colhead{}              &
\colhead{start}         &
\colhead{end}           &
\colhead{}              &
\colhead{}              &
\colhead{}              &
\colhead{}              &
\colhead{}              &
\colhead{}              &
\colhead{}              \\
\colhead{}              &
\colhead{day}           &
\colhead{day}           &
\colhead{s}             &
\colhead{ct\,ks$^{-1}$}  &
\colhead{ct\,ks$^{-1}$}  &
\colhead{ct\,ks$^{-1}$}  &
\colhead{ct\,ks$^{-1}$}  &
\colhead{}              &
\colhead{erg\,s$^{-1}$}              
}
\startdata
PTF09atu   &$    2.7$&  \nodata  &     4858.0&$   -0.02$&\nodata &\nodata &     0.62& 1.00&$   <1.9\times10^{44}$\\
PTF09cnd   &$  -17.6$&  \nodata  &     3441.8&$   -0.02$&\nodata &\nodata &     0.87& 1.00&$   <5.4\times10^{43}$\\
           &$  -13.7$&  \nodata  &     3557.2&$   -0.01$&\nodata &\nodata &     0.84& 1.00&$   <5.3\times10^{43}$\\
           &$  -10.5$&  \nodata  &     3273.1&$   -0.02$&\nodata &\nodata &     0.92& 1.00&$   <5.7\times10^{43}$\\
           &$   -5.9$&  \nodata  &     3997.8&$   -0.02$&\nodata &\nodata &     0.75& 1.00&$   <4.7\times10^{43}$\\
           &$   -2.2$&  \nodata  &     2232.1&$   -0.05$&\nodata &\nodata &     1.34& 1.00&$   <8.4\times10^{43}$\\
           &$    4.5$&  \nodata  &     2980.6&$   -0.03$&\nodata &\nodata &     1.01& 1.00&$   <6.3\times10^{43}$\\
           &$   18.3$&  \nodata  &     2026.8&$   -0.01$&\nodata &\nodata &     1.48& 1.00&$   <9.2\times10^{43}$\\
           &$   27.9$&  \nodata  &     1910.6&$   -0.02$&\nodata &\nodata &     1.57& 1.00&$   <9.8\times10^{43}$\\
           &$  -17.6$&$    -2.2 $&    16501.9&$   -0.02$&\nodata &\nodata &     0.18& 1.00&$   <1.1\times10^{43}$\\
           &$    4.5$&$    27.9 $&     6917.9&$   -0.02$&\nodata &\nodata &     0.43& 1.00&$   <2.7\times10^{43}$\\
           &$  -17.6$&$    27.9 $&    23419.8&$   -0.02$&\nodata &\nodata &     0.13& 1.00&$   <8.0\times10^{42}$
\enddata
\tablecomments{Summary of all the 305 {\it Swift}-XRT flux measurements
of the 28 SNe in our sample.
For each SN we list the observation date ($t-t_{{\rm peak}}$) relative
to the $t_{{\rm peak}}$ listed in Table~\ref{tab:Samp}.
Rows that have both a start day and end day listed,
give 'super-epoch' measurements as described in the main text.
Exp. is the exposure time, CR, $\Delta$CR$_{-}$, $\Delta$CR$_{+}$
are the source counts rate, lower error and upper error, respectively.
CR$_{{\rm UL}}$ is the 2-$\sigma$ upper limit on the source count rate,
FAP is the false alarm probability (see text),
and $L_{{\rm X}}$ is the source luminosity,
or the the 2-$\sigma$ upper limit on the luminosity,
in the 0.2--10\,keV band.
If FAP$>0.01$ than we provide the 2-$\sigma$ upper limits,
otherwise the measurements are given along with the errors.
This table is published in its entirety in the electronic edition of
the {\it Astrophysical Journal}. A portion of the full table is shown here for
guidance regarding its form and content.}
\label{tab:Obs}
\end{deluxetable*}

Several objects listed in Table~\ref{tab:Obs}
show count rates which may deviate from zero.
Here we discuss the observations of all seven sources that 
have FAP$\le0.01$ in at least one of the epochs or super-epochs.
We note that Table~\ref{tab:Obs} contains 305 epochs and super-epochs.
Therefore, we expect about three random false detections.
Interpretation of these observations is discussed in \S\ref{sec:obsmoel}.

\begin{figure}
\centerline{\includegraphics[width=8.5cm]{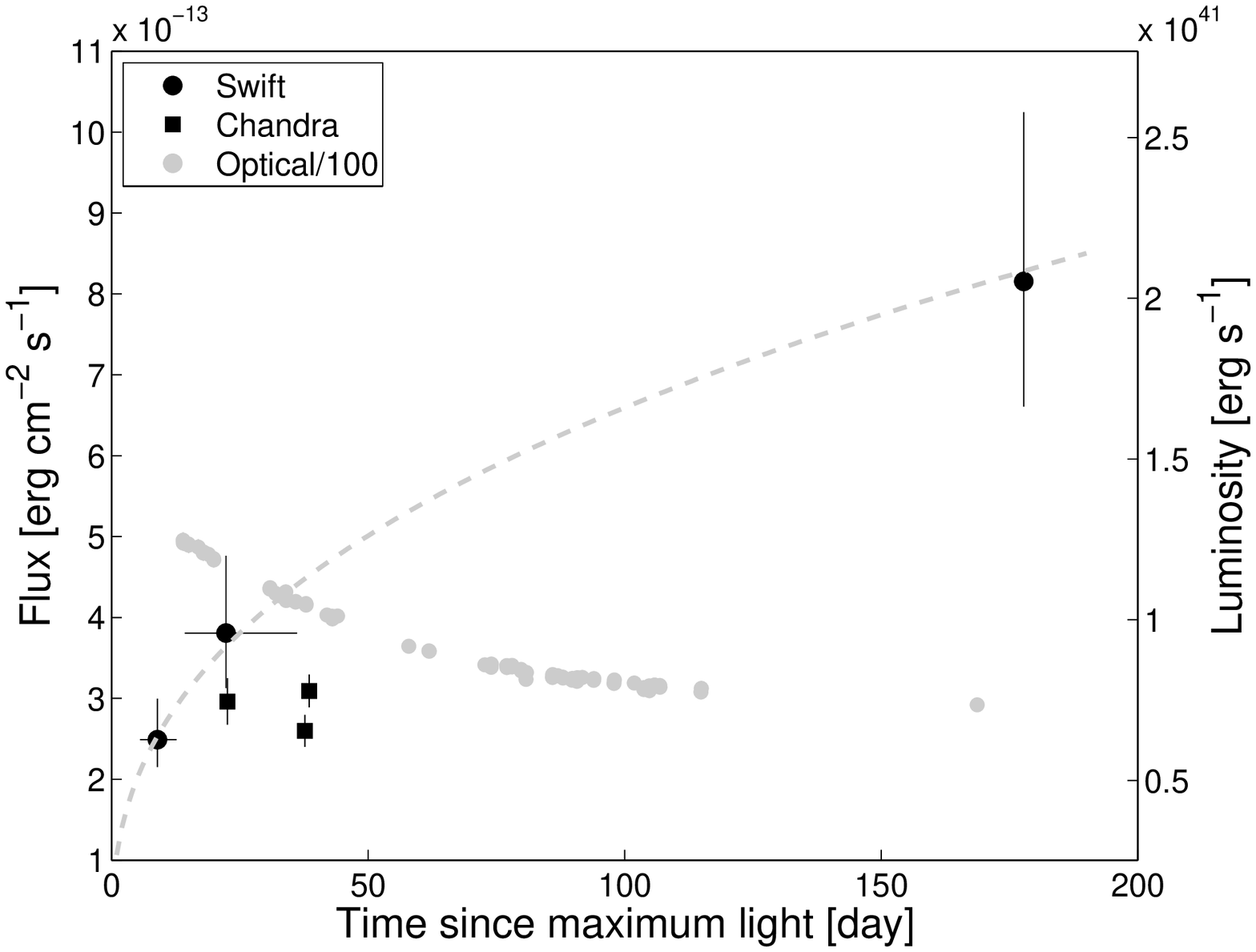}}
\caption{The {\it Swift} (circles) and {\it Chandra} (squares)
X-ray light curve extracted at the position of SN\,2010jl.
The unabsorbed flux was calculated using
PIMMS
assuming Galactic column density of $N_{{\rm H}}=3.05\times10^{20}$\,cm$^{-2}$,
and $N_{{\rm ph}}(E)\propto E^{-0.2}$ power-law spectrum.
We note that the {\it Chandra} observations show
a possible extended source near the SN location.
This additional source may contaminate the {\it Swift}-XRT
measurements and can explain the small discrepancy
between {\it Chandra} and {\it Swift}-XRT.
Alternatively, the discrepancy between the {\it Chandra} and {\it Swift}
light curves can be explained if the X-ray spectrum is
harder or the $N_{{\rm H}}$ is larger than we assumed.
We note that for N$_{{\rm H}}$ which is 1000 larger than
the Galactic value, the unabsorbed {\it Swift} 
({\it Chandra}) flux will be about 5.2 (7.2) times larger.
For reference, the grey circles show the PTF $R$-band luminosity
of this SN scaled by 0.01.
The PTF $R$-band luminosity was calibrated using
the method described in Ofek et al.\ (2012a) and calibration
stars listed in Ofek et al.\ (2012b).
\label{fig:PTF10aaxf_LCX}}
\end{figure}

{\it SN\,2010jl/PTF\,10aaxf} (Fig.~\ref{fig:PTF10aaxf_LCX}):
This SN has a large number of {\it Swift}-XRT
observations, as well as {\it Chandra} ACIS-S observations in five epochs
(Chandra et al.\ 2012a),
of which three are public (PIs: Pooley; Tremonti).
The host, SDSS\,J094253.43$+$092941.9,
is an irregular star-forming galaxy.
The Binned {\it Swift}-XRT and {\it Chandra} light curves,
as well as the PTF $R$-band light curve, are
presented in Figure~\ref{fig:PTF10aaxf_LCX}.

\begin{figure}
\centerline{\includegraphics[width=8.5cm]{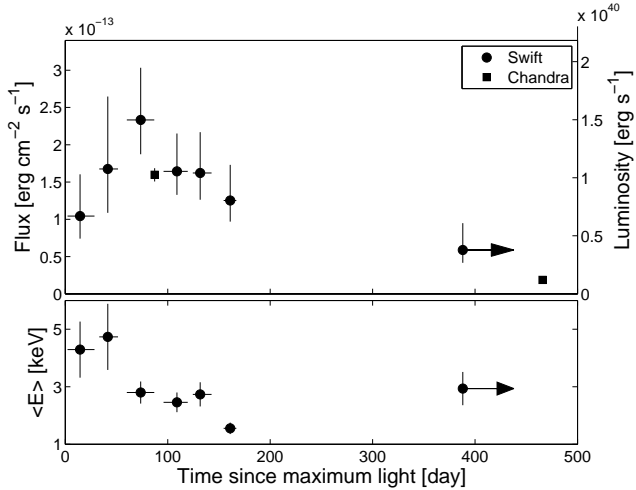}}
\caption{{\it Upper panel}:
The {\it Swift} (circles) and {\it Chandra} (squares)
X-ray light curve extracted at the position of SN\,2006jc.
The unabsorbed flux was calculated using
PIMMS
assuming Galactic column density of $N_{{\rm H}}=1.0\times10^{20}$\,cm$^{-2}$,
and $N_{{\rm ph}}(E)\propto E^{-0.2}$ power-law spectrum.
{\it lower panel}: The mean photon energy of the {\it Swift}-XRT
X-ray observations in the 0.2--10\,keV band as a function of time.
\label{fig:SN2006jc_LCX}}
\end{figure}

\begin{figure}
\centerline{\includegraphics[width=8.5cm]{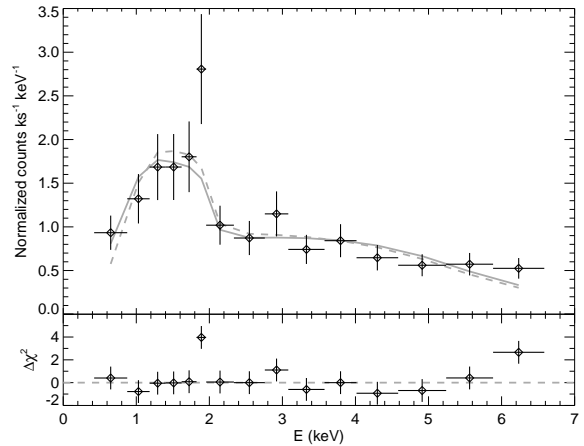}}
\caption{
spectrum of SN\,2006jc, as observed at 87 days after maximum optical light,
near or at peak X-ray luminosity. The top panel shows the binned X-ray data
(points) with errors (1$\sigma$) along with the best-fit model,
a photon index $\Gamma=0.24^{+0.22}_{-0.16}$ (90\%-confidence) power-law,
where $N_{{\rm ph}}\propto E^{-\Gamma}$,
and with negligible line of sight absorption.
The dashed line shows the 
constrained best-fit that has maximal
$N_{{\rm H}}$ (90\%-confidence; $1.37\times10^{21}$\,cm$^{-2}$). 
The bottom panel shows the $\Delta\chi^2$ residuals
to the best fit. The fit is acceptable, with $\chi^{2}=11.75$ for 12 degrees
of freedom ($p_{\rm null}=0.47$).
\label{fig:SN2006jc_XraySpec}}
\end{figure}

{\it SN\,2006jc} (Fig.~\ref{fig:SN2006jc_LCX}):
this is the only SN in our sample
that belongs to the rare class of type-Ibn SNe.
This SN has a large number of {\it Swift}-XRT
and {\it Chandra} observations.
The SN is detected
on multiple epochs and its X-ray light curve 
is presented in Figure~\ref{fig:SN2006jc_LCX} (see also Immler et al.\ 2008).
The SN was detected in X-rays soon after maximum optical light
and reached a maximum X-ray luminosity of about
$1.5\times10^{40}$\,erg\,s$^{-1}$ at $\approx100$\,days
after maximum optical light,
$\gtorder 6$ times the SN rise time.
SN\,2006jc was observed by {\it Chandra}
on several occasions.
We reduced a 55\,ks {\it Chandra} observation
with 306 photons at the SN location
taken 87 days after maximum light.
Given the limited number of photons we did~not
attempt to fit complex models.
We found that the spectrum is well fitted by
an $N_{{\rm ph}}(E)\propto E^{-0.2}$ power-law and with
negligible absorbing column density.
The spectra and the best fit model are presented in
Figure~\ref{fig:SN2006jc_XraySpec}.
Regardless of the exact spectral shape, the spectrum
is hard.
Marginalizing over all the free parameters,
we find a 2-$\sigma$ upper limit of
$N_{{\rm H}}<1.26\times10^{21}$\,cm$^{-2}$,
in excess of the Galactic column density.

\begin{figure}
\centerline{\includegraphics[width=8.5cm]{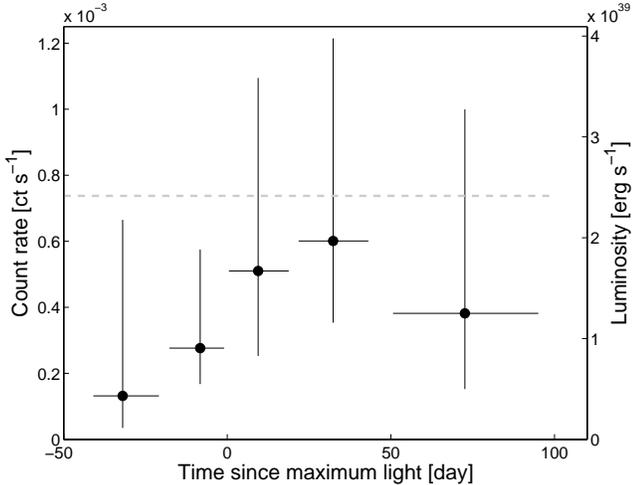}}
\caption{The unabsorbed {\it Swift}-XRT light curve extracted at the
position SN\,2011ht in the 0.2--10\,keV band, corrected
for the aperture size and assuming $N_{{\rm H}}=7.8\times10^{19}$\,cm$^{-2}$.
The dashed gray line represents the 2-$\sigma$ upper limit on the flux
from two combined {\it Swift}-XRT observations obtained
$-1649$ and $-1405$ days prior to maximum light. 
\label{fig:SN2011ht_LCX}}
\end{figure}

{\it SN\,2011ht} (Fig.~\ref{fig:SN2011ht_LCX}): 
The SN took place about $21''$ from the center of UGC\,5460.
It was
observed on multiple epochs using {\it Swift}-XRT.
Roming et al.\ (2012) reported
a detection of an X-ray source at the SN position.
The binned {\it Swift}-XRT X-ray light curve of this SN is shown in
Figure~\ref{fig:SN2011ht_LCX}.
Apparently the light curve is rising, peaking
at around 40\,day after maximum optical light,
and then declines.
However, the uncertainties in the flux measurements are large
and the light curve is consistent with being flat
(i.e., a best fit flat model gives $\chi^{2}/dof=1.23/4$).
Moreover, recently Pooley (2012)
reported on a 9.8\,ks {\it Chandra}
observation\footnote{At the time of the writing this paper, this observation
was proprietary.}
of this SN.
He argued that the emission detected by Roming et al.\ (2012)
which is shown in Figure~\ref{fig:SN2011ht_LCX}
is from a nearby source found 4.7'' from the SN location.

%

\begin{figure}
\centerline{\includegraphics[width=8.5cm]{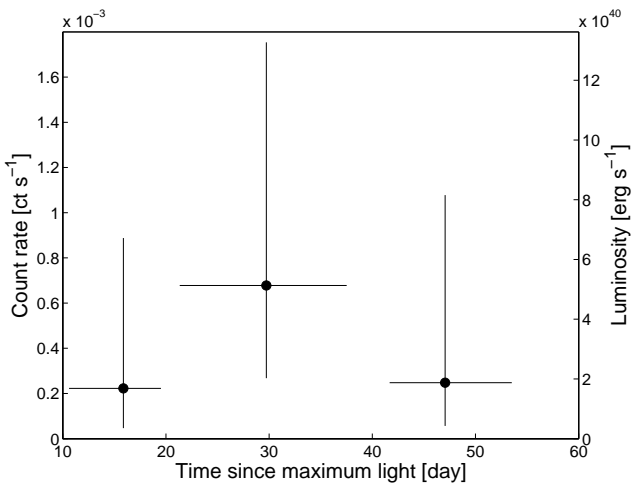}}
\caption{The unabsorbed {\it Swift}-XRT light curve extracted at the
position SN\,2010al in the 0.2--10\,keV band, corrected
for the aperture size and assuming $N_{{\rm H}}=3.92\times10^{20}$\,cm$^{-2}$.
\label{fig:SN2010al_LCX}}
\end{figure}

{\it SN\,2010al} (Fig.~\ref{fig:SN2010al_LCX}): 
This SN was found $12''$ from the center of the spiral galaxy UGC\,4286.
The SN was observed in multiple epochs using {\it Swift}-XRT
with a total integration time of 35\,ks.
It is detected in the combined image with a mean luminosity
of $\approx7\times10^{39}$\,erg\,s$^{-1}$.
Figure~\ref{fig:SN2010al_LCX} presents the binned
{\it Swift}-XRT light curve of this SN.
Although the light curve peaks around 30\,days after maximum light,
it is consistent with being flat
(i.e., a best fit flat model gives $\chi^{2}/dof=0.15/2$).
%

{\it SN\,2005db}: This SN was observed on three epochs,
about two years after maximum light (676 to 695\,days),
using {\it Swift}-XRT.
The combined image, with an exposure time of 13.6\,ks,
shows a faint source (5 photons)
with a false alarm probability of $2\times10^{-4}$ per trial.
However, given the fact that we have 305 observations,
the false alarm probability over all trials
is about $0.06$.
Using the Galactic column density (Table~\ref{tab:Samp})
and assuming a power-law spectrum $N_{{\rm ph}}(E)\propto E^{-0.2}$,
the unabsorbed flux is 
$(1.97_{-0.72}^{+1.7})\times10^{40}$\,erg\,s$^{-1}$.
{\it Chandra} observed this target twice at
722.7 and 1051.4\,days after maximum light (PI: Pooley),
with integrations of 3.0 and 5.0\,ks, respectively.
Using the same assumptions as above, we put a
$2$-$\sigma$ upper limit on the unabsorbed flux of
$1.4\times10^{40}$\,erg\,s$^{-1}$ for both epochs.
To conclude, given the uncertainties, the {\it Chandra} upper limits
are consistent with the possible {\it Swift} detection.
However, given that this source has a single
detection we cannot firmly conclude
that the detection is real.

{\it PTF\,11iqb}: This SN has multi-epoch {\it Swift}-XRT observations
taken between about $-13$ and 28\,days relative to maximum light.
The SN is detected in a single epoch about 24\,days
after maximum light, with a false alarm probability of $1.4\times10^{-3}$
per trial.
However, given that we are reporting 305
individual X-ray observations, the false alarm
probability over all the trials is $0.45$.
The SN was not detected at the last epoch, 28 days
after maximum light.
Unfortunately, this object has no further {\it Swift}-XRT observations.

\begin{figure}
\centerline{\includegraphics[width=8.5cm]{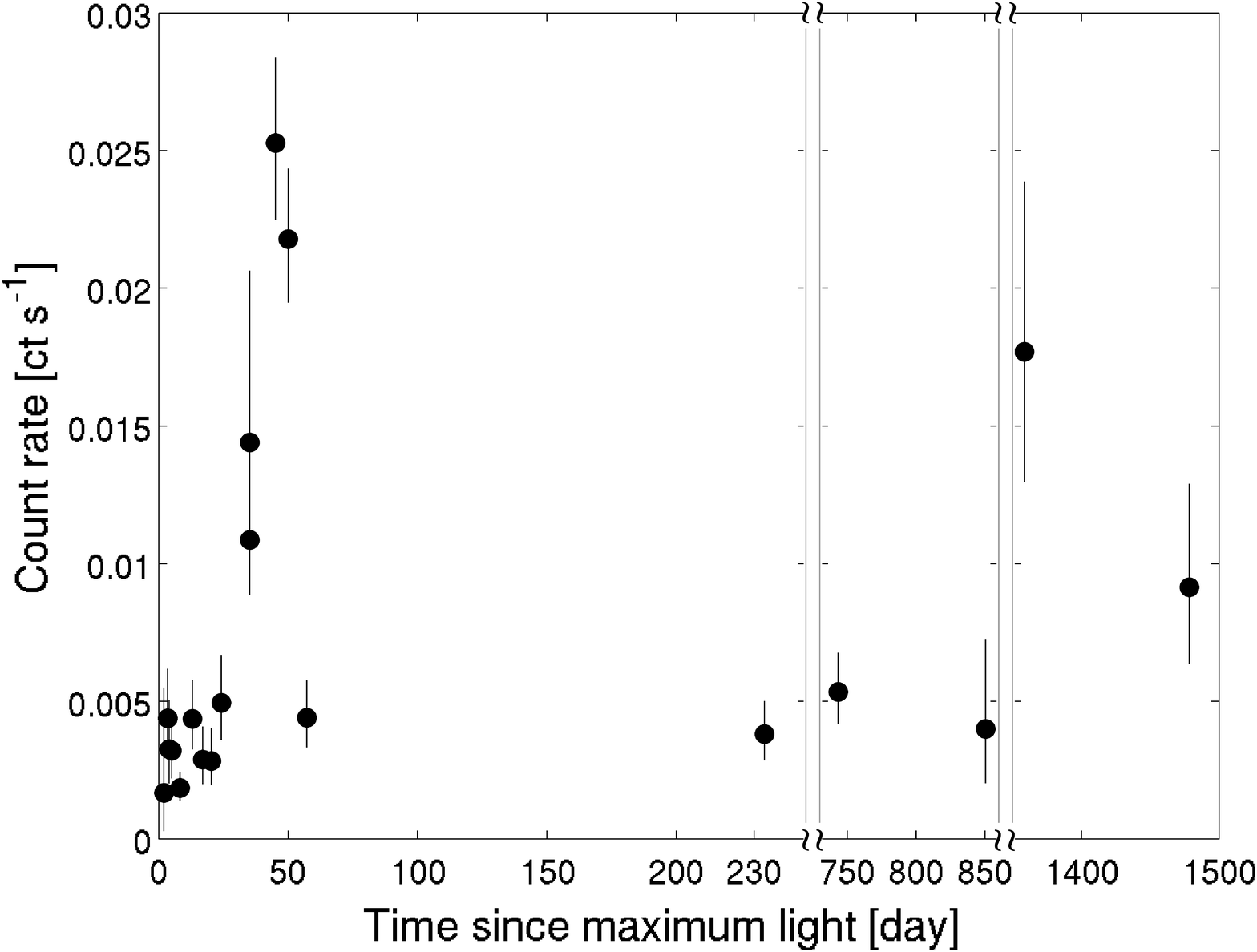}}
\caption{X-ray light curve, corrected for the aperture size,
extracted at the
position SN\,2007pk.
The emission, and flare, are likely due to AGN activity in
the host galaxy NGC\,579.
\label{fig:SN2007pk_LCX}}
\end{figure}

{\it SN\,2007pk} (Fig.~\ref{fig:SN2007pk_LCX}):
This SN has a large number of {\it Swift}-XRT
observations,
as well as a {\it Chandra} observation (PI: Pooley).
Immler et al.\ (2007) reported a tentative detection
of this SN in the images taken between MJD 54417.09 and 54420.04.
The light curve of the source extracted in the SN
position is shown in Figure~\ref{fig:SN2007pk_LCX}.
The light curve shows a brightening, with a peak around MJD 54461,
and a full width at half maximum of about 40\,days.
However, the SN is about $7''$
from the center of the spiral host galaxy,
NGC\,579, and
the centroids of the {\it Swift}-XRT positions in
individual exposures
are clustered around the galaxy nucleus, rather than the SN position.
We note that emission from the center
of NGC\,579 is clearly detected in the {\it Chandra} observation
and that there is some emission in the source position.
However, this emission may be due to diffuse emission
from NGC\,579.
Therefore, without conclusive evidence that the emission
is from the SN, here we assume that the observed flare
as well as the quiescent X-ray emission from
the position of the source is due to AGN activity
in NGC\,579.
In Table~\ref{tab:Samp}, we adopted an upper limit on
the X-ray luminosity of SN\,2007pk, which is based on
the average luminosity observed from the direction of the source,
presumably due to AGN activity.

\section{The model}
\label{sec:Disc}

Several recent works discuss the possibility of detecting X-ray
emission from
SN collisionless shocks in wind environments
(Katz et al.\ 2011; Murase et al. 2011;
Chevalier \& Irwin 2012; Svirski et al.\ 2012).
Here, we review the main processes relevant to shock
breakout in wind-profile CSM (\S\ref{sec:cond}),
emission from collisionless shocks,
including the importance of
bound-free absorption (\S\ref{sec:coll}),
and the possibility of detecting radio emission (\S\ref{sec:radio}).
In \S\ref{sec:obsmoel} we discuss our observations
in the context of this model.

\subsection{Shock breakout conditions in wind-profile CSM}
\label{sec:cond}

Here, we assume that the CSM around the progenitor has wind-density profile
$\rho = Kr^{-2}$, where $K\equiv\dot{M}/(4\pi v_{w})$
is the mass-loading parameter,
$\dot{M}$ is the progenitor mass-loss rate, and $v_{w}$ is the
progenitor wind speed.
The Thomson optical depth, $\tau$,
due to an ionized progenitor wind between the observer and
a spherical surface at radius
$r$ from the star center is
\begin{eqnarray}
\tau  & \approx & \frac{\kappa \dot{M}}{4\pi v_{{\rm w}}r} \cr
      & \cong   & 170 \kappa_{0.34} \dot{M}_{0.1} v_{{\rm w},10}^{-1} r_{15}^{-1}.
\label{tau}
\end{eqnarray}
Here $\kappa_{0.34}$ is the opacity in units of 0.34\,cm$^{2}$\,gr$^{-1}$,
$\dot{M}_{0.1}$ is the mass-loss rate in units of 0.1\,M$_{\odot}$\,yr$^{-1}$,
$v_{{\rm w},10}$ is the wind speed in units of 10\,km\,s$^{-1}$,
and $r_{15}$ is the radius in units of $10^{15}$\,cm.
We note that this relation is correct up to a 
correction factor of the order of unity (see Balberg \& Loeb 2011).
The photons in the radiation dominated and mediated shock
from the SN explosion
breakout
when $\tau\approx c/v_{{\rm s}}$ (e.g., Weaver 1976; Ofek et al.\ 2010),
where $c$ is the speed of light and $v_{{\rm s}}$ is the shock velocity.
At this optical depth, the photon diffusion time becomes shorter
than the hydrodynamical time scale (i.e., $r/v_{{\rm s}}$)
and the photons can diffuse outward faster than the ejecta.
Therefore, the condition for the shock breakout
to take place in a steady wind environment ($w=2$) is
\begin{eqnarray}
\frac{\dot{M}_{0.1}}{v_{{\rm w},10}} & \gtorder & 1.2\times10^{-6} \tau_{30} r_{\odot} \kappa_{0.34}^{-1}\,(0.01\,{\rm M}_{\odot}\,{\rm yr}^{-1}\,{\rm km}^{-1}\,{\rm s}),
\label{Mv}
\end{eqnarray}
where $\tau_{30}$ is the Thomson optical depth in units of 30,
and $r_{\odot}$ is the radius in units of the solar radius.

\begin{figure}
\centerline{\includegraphics[width=8.5cm]{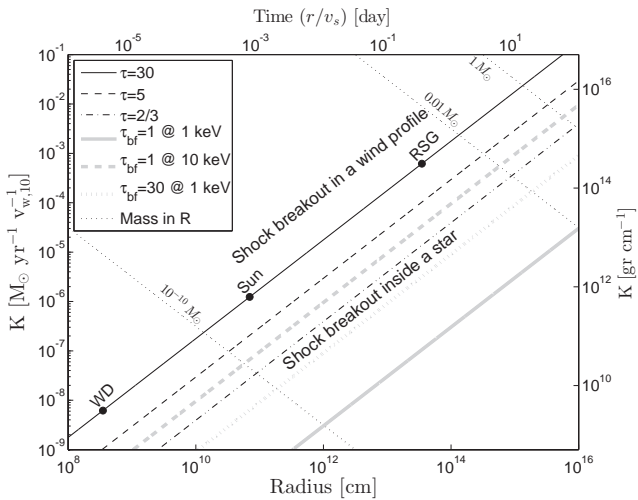}}
\caption{Various regions in the radius vs.\ wind mass-loading parameter
phase space, assuming $w=2$.
As indicated in the legend
the black-solid line represent Thomson optical depth, $\tau=30$
(i.e., shock breakout with $v_{{\rm s}}=10^{4}$\,km\,s$^{-1}$).
$\tau=5$ and $\tau=2/3$ are marked with the black dashed,
and black dashed-dotted lines, respectively.
The dotted black lines represent the integral of mass
inside a given radius (i.e., $\int_{0}^{r}{4\pi r^{2} Kr^{-2}dr}=4\pi K r$).
The gray lines show constant bound-free absorption
at a given energy.
For reference, 
the radii of red supergiant (RSG; 500\,R$_{\odot}$),
the Sun and a massive white dwarf (WD; 0.005\,R$_{\odot}$)
are marked in circles on the $\tau=30$ line.
See text for discussion.
\label{fig:Radius_K}}
\end{figure}

Figure~\ref{fig:Radius_K} shows the radius vs.\ the mass-loading parameter
at which $\tau\approx 30$ (Eq.~\ref{Mv}; i.e., $v_{{\rm s}}=10^{4}$\,km\,s$^{-1}$; black-solid line).
For example, this plot suggests that the critical mass-loading,
above which the shock will breakout in the wind environment,
is $\sim 6\times10^{-4}$\,M$_{\odot}$\,yr$^{-1}$\,v$_{{\rm w},10}^{-1}$
for a 500\,R$_{\odot}$ red supergiant.
Assuming $v_{{\rm s}}=10^{4}$\,km\,s$^{-1}$, systems found above
the solid line will have a shock breakout in a wind profile
(the two cases discussed in Chevalier \& Irwin 2011).
Objects found in regions below the
$\tau\approx2/3$, black dashed-dotted line, will have
a shock breakout within the stellar surface and the
wind will~not influence the diffusion of energy
from the shock breakout.
Finally, systems below the solid line ($\tau\approx30$)
and above the dashed-dotted line ($\tau\approx2/3$) will have a shock breakout
below the stellar surface, but the wind can play a 
role in the diffusion of the shock energy (e.g., Nakar \& Sari 2010).

We note that, in order to form a type-IIn SN,
we require optically thin material,
which is ionized by the SN radiation field.
Therefore, we speculate that type-IIn SNe can be found
below and above the $\tau\approx2/3$ line,
and that not all type-IIn SNe are powered
by shock breakout in dense CSM environments.

\subsection{X-ray emission from collisionless shocks}
\label{sec:coll}

Katz et al.\ (2011) showed that if the shock width ($\Delta r$)
at radius $r$
is of the order of $r$
(i.e., the Thomson optical depth vary on scales of the order
of the radius $r$),
the radiation mediated and dominated shock
will transform into a collisionless shock,
and hard X-ray photons will be generated.
It is reasonable to assume that some of this energy will be
produced in the 1--10\,keV X-ray regime.
We note that the exact spectrum was~not calculated self-consistently
and, therefore, is not known.

Chevalier \& Irwin (2012) and Svirski et al.\ (2012)
showed that, during the first several shock breakout time scales
after the shock breakout,
the optical depth is too large for the hard X-rays to escape.
They found that the most efficient processes 
in blocking the hard X-rays ($\sim100$\,keV) are likely:
(i) cooling of the electrons by inverse Compton scattering
on soft photons generated by the pre-shocked material;
and (ii) Compton scattering.
On average, for each Compton scattering, the photon
loses a fraction of energy which is comparable to
$4k_{{\rm B}}T_{{\rm e}}/(m_{{\rm e}}c^{2})$
(e.g., Lang 1999).
Here $k_{{\rm B}}$ is the Boltzmann constant, $T_{{\rm e}}$ is the electron
temperature, and $m_{{\rm e}}$ is the electron mass.
Svirski et al.\ (2012) argued that when the Thomson optical
depth declining to $\tau\approx 5$,
the hard X-ray emission ($\sim 100$\,keV) will be Comptonized
to the 1--10\,keV X-ray band and may diffuse
out of the CSM.
Since the optical depth in a wind profile ($w=2$) 
decrease as $r^{-1}$ (Equation~\ref{tau}),
the collisionless shock signature will be observable in the X-ray regime
only when the optical depth decreases by an order of magnitude
(i.e., the shock radius increases by an order of magnitude).
Given that the shock velocity falls as $t^{-1/5}$,
they argued that this should happen roughly between 10 to 50 times
the shock-breakout time scale.
For reference, we show the $\tau=5$ line in Figure~\ref{fig:Radius_K}
(dashed line; assuming $v_{{\rm s}}=10^{4}$\,km\,s$^{-1}$).
Such late-time emission may be relevant only
if the collisionless shock is still important
at late times and if the Comptonization
remains the dominant process.

Chevalier \& Irwin (2012) consider the effect of bound-free
absorption. Since, above 0.5\,keV, metals with a high ionization potential
dominate the bound-free absorption,
full ionization is required in order to avoid 
absorption by this process.
They argued that an ionization
parameter of $\sim1000$ is needed to achieve full ionization
(including the metal atoms' inner electrons).
They estimate that full ionization is achieved
for shocks with velocity $v_{{\rm s}}\gtorder10^{4}$\,km\,s$^{-1}$.
In order to estimate the effect of bound-free absorption, we need
to evaluate the density of the CSM.

Assuming a hydrogen-rich material, the particle density in a wind profile is given by
\begin{eqnarray}
n      & =       & \frac{1}{\langle \mu_{{\rm p}}\rangle} \frac{\dot{M}}{4\pi m_{{\rm p}} v_{{\rm w}}r^{2}} \cr
       & \cong   & 3.02\times10^{11} \frac{1}{\langle \mu_{{\rm p}}\rangle} \dot{M}_{0.1} v_{{\rm w},10}^{-1} r_{15}^{-2}\,{\rm cm}^{-3},
\label{n}
\end{eqnarray}
where 
$\langle \mu_{{\rm p}}\rangle$ is the mean number of protons per particle
(mean molecular weight).
For our order of magnitude calculation, we assume
$\langle \mu_{{\rm p}}\rangle=1$.
In a wind profile, the column density between the radius $r$ and the observer is:
\begin{equation}
N      \sim 3.02\times10^{26} \dot{M}_{0.1} v_{{\rm w},10}^{-1} r_{15}^{-1}\,{\rm cm}^{-2}.
\label{N}
\end{equation}
Assuming the gas in the pre-shocked wind is neutral,
the bound-free optical depth in the 0.03 to 10\,keV region
is roughly given by
(e.g., Behar et al.\ 2011)\footnote{This approximation deviates by a factor of two from a more accurate calculation (e.g., Morrison \& McCammon 1983).}:
\begin{eqnarray}
\tau_{{\rm bf}} & =       &  N\sigma(E) \cr
             & \sim & 3\times10^{4} \dot{M}_{0.1} v_{{\rm w},10}^{-1} r_{15}^{-1} E_{1}^{-2.5},
\label{tauX}
\end{eqnarray}
where $\sigma(E)$ is the bound-free cross section as a function
of energy $E$, and $E_{1}$ is the energy in keV.
This approximation is valid when the material
is neutral. However, since above $\sim0.5$\,keV,
metals with a high ionization potential
dominate the absorption,
this formula is still valid, to an order of a magnitude,
above 0.5\,keV when some (or even one) of the inner electrons
of the metals are bound.
In Figure~\ref{fig:Radius_K} we show the lines
at which $\tau_{{\rm bf}}\approx1$ at 1\,keV and 10\,keV,
and at which $\tau_{{\rm bf}}\approx30$ at 1\,keV.
Comparing equations~\ref{tau}
and \ref{tauX} suggests that at the time of shock breakout
the bound-free cross-section in the $\approx 0.5$--$8$\,keV
range is larger than the Thomson cross section.
We note that this may modify the properties of the shock breakout
and its spectrum.
Moreover, the $\tau_{{\rm bf}}=1$ line at 1\,keV is located far
below the $\tau=5$ line.
This suggests
that if the pre-shocked wind is not completely ionized
(e.g., $v_{{\rm s}}\ltorder 10^{4}$\,km\,s$^{-1}$),
then soft ($\ltorder 1$\,keV) X-ray emission is not
expected in the simple case of a spherically symmetric wind ($w=2$) profile,
even at late times.
Moreover, the $\tau_{{\rm bf}}=1$ at 10\,keV line is located slightly below
the $\tau=5$ line.
Therefore, bound-free absorption is likely important
even at late stages.
This may indicate
that $\sim$10\,keV X-rays may
escape the wind on a time scale somewhat longer than suggested by
Svirski et al.\ (2012),
and that observations at energies above 10\,keV
may be more effective (e.g., by the {\it NuStar} mission;
Harrison et al.\ 2010).

All the order of magnitude calculations presented so far assume a wind
density profile with $w=2$.
However, we note that if the CSM profile falls
faster than $r^{-2}$,
or alternatively if the wind is not spherically symmetric or is clumpy,
then the column density may fall faster than $r^{-1}$
in some directions (e.g., Eq.~\ref{N}),
and it will enable the X-rays to escape earlier
than predicted by Svirski et al.\ (2012)
for the $w=2$ case.

\subsection{Radio emission from collisionless shocks}
\label{sec:radio}

The shock going through the CSM may generate
synchrotron radiation peaking at radio frequencies
(e.g., Slysh 1990;
Chevalier \& Fransson 1994;
Chevalier 1998;
Horesh et al.\ 2012;
Krauss et al.\ 2012;
Katz 2012).
However, if the material is ionized or partially ionized
then the free-free optical-depth may block
this radiation.
In order to test if radio signature is expected,
we need to estimate the free-free optical depth
(e.g., Chevalier 1981)
which, for a ionized CSM with a wind profile,
is given by (Lang 1999, Equation 1.223)
\begin{eqnarray}
\tau_{{\rm ff}} & \approx & 2.6\times10^{10} T_{{\rm e}, 4}^{-1.35} \nu_{10}^{-2.1} v_{{\rm w},10}^{-2} \dot{M}_{0.1}^{2} r_{15}^{-3} \cr
              & \propto & r_{15}^{1-2w},
\label{tau_ff}
\end{eqnarray}
where $\nu_{10}$ is the frequency in units of 10\,GHz.
The free-free optical depth is so large that radio emission is not expected.
However, if in some regions
the CSM density profile falls
significantly faster than $r^{-2}$
then the free-free absorption may be low enough
for radio emission (e.g., synchrotron)
to escape the CSM.
We predict that if the CSM is ionized,
then due to the effect of free-free absorption,
the synchrotron radio emission generated
in the shocked CSM may have a relatively steep
radio spectrum.
Therefore, it is preferable to search for
this emission at high frequencies and late times
after maximum light.
However, the existence of radio emission is likely very
sensitive to the exact properties, like density profile,
symmetry, and homogeneity of the CSM.
We speculate that good candidates for radio emission
will be SNe in which the wind filling factor is low, or asymmetric,
or alternatively when the wind is ejected in a relatively short eruption,
and therefore may have a steep density profile (see \S\ref{sec:obsmoel}).

We note that van Dyk et al.\ (1996) presented a search for radio
emission among ten type-IIn SNe.
None of the SNe in their sample was detected in radio.
However, these observations were conducted between 2\,yr to 14\,yr
after the SN explosion.

\section{The observations in context of the model}
\label{sec:obsmoel}

As shown in Table~\ref{tab:Obs} and
Figures~\ref{fig:PTF10aaxf_LCX}--\ref{fig:SN2010al_LCX},
several SNe in our sample show X-ray emission in the 0.2--10\,keV
range.
Some SNe are presumably detected (and maybe peaking)
in X-rays near maximum optical light
(i.e., SN\,2011ht; SN\,2010al).
However, the X-ray luminosity of SN\,2011ht and SN\,2010al
at maximum visible light
is about $0.05$ and $0.3$, respectively,
of their visible light luminosity.
These X-ray luminosities are higher
than predicted by Svirski et al.\ (2012; i.e. X-rays $10^{-4}$ of optical)
for the CSM shock breakout case.
Moreover, although the X-ray light curves are consistent
with a non-variable luminosity, it is
possible that they are peaking near maximum optical light.
Therefore, we conclude that the optical
light curves of SN\,2011ht and SN\,2010al,
which are type-IIn SNe, are likely not
powered by a shock breakout in CSM.

SN\,2010jl (Chandra et al.\ 2012a) and SN\,2006jc 
(Immler et al.\ 2007) are detected in X-rays
and are peaking at late times $\gtorder 10 t_{{\rm rise}}$
and $\gtorder 6 t_{{\rm rise}}$, respectively.
Moreover, the X-ray luminosity at maximum
visible light is about $10^{-3}$ of the visible-light
luminosity.
This is roughly consistent with the predictions
of Svisrski et al.\ (2012).
Therefore, we conclude that these two SNe
are likely powered by shock breakout in CSM.
We discuss these SNe in detail in \S\ref{SN2010jl}-\ref{SN2006jc}.

Two other SNe in our sample,
PTF\,11iqb and SN\,2005db,
have marginal X-ray detections. 
Therefore, we cannot make any firm
conclusion regarding the reality and nature of this emission.

As indicated in the column $L_{X}/L_{{\rm opt}}$ in
Table~\ref{tab:Samp}, the rest of the SNe in our sample,
including all the SLSN-I events, do~not have
late time observations and/or deep 
enough limits in order to evaluate their nature.
Our upper limits are mostly obtained at early times
($\ltorder 3t_{{\rm rise}}$)
after the SN explosion,
or at very late times ($\gtorder 37t_{{\rm rise}}$).
We note that all the observations of the
hydrogen-poor luminous SNe were obtained at $\ltorder 2t_{{\rm rise}}$
after optical maximum light.

We note that recently Chandra et al.\ (2012b)
reported on X-ray and radio observations of another
type-IIn event, SN\,2006jd.
This event is listed as a type-IIb SNe in the IAUC
SN list\footnote{http://www.cbat.eps.harvard.edu/lists/Supernovae.html}
and, therefore, was~not included in our sample.
We note that the X-ray observations of this event
started about a year after the explosion, so there is no
measurement of $L_{{\rm X}}/L_{{\rm opt}}$ during optical maximum light.

We conclude that deeper X-ray observations
over longer periods of time, since maximum optical light,
are required in order to understand the nature of type-IIn and SLSN-I 
events.
The current null detection of hydrogen-poor
luminous SNe in X-rays
cannot be used to reject the CSM-interaction model
proposed by Quimby et al.\ (2011a).

\subsection{SN\,2010jl (PTF\,10aaxf)}
\label{SN2010jl}

The X-ray luminosity of SN\,2010jl near optical maximum light is about
$6\times10^{40}$\,erg\,s$^{-1}$,
which is $\sim10^{-3}$ of its $R$-band luminosity.
For the case of shock breakout in CSM,
Svirski et al.\ (2012) predicted that
near optical maximum light the X-ray luminosity will be
about $10^{-4}$ of the optical luminosity.
However, it is possible that the bolometric optical luminosity is
higher (e.g., due to metal blanketing).
Moreover, we note that for $w>2$, the
amount of material above the shock is smaller and
the X-rays will be less effected by absorption at early times.

The {\it Swift} X-ray flux is rising with time,
and is consistent 
with a power-law of the form $\approx t^{0.4}$,
where $t$ is the time since optical maximum light.
Svirski et al.\ (2012) predicted that the hard radiation
(i.e., X-ray) component
of a collisionless shock will rise as $\propto t^{\beta}$
with $\beta$ is between $0$ to $2$
when the ejecta are colliding with a wind mass that
is comparable to its own mass,
and with $\beta$ between $1$ to $5/2$
when the wind is less massive than the ejecta.
However, our observations constrain only
the $0.2$--$10$\,keV range, while the hard component
can emit at energies up to $\sim100$\,keV.
Moreover, Svirski et al.\ (2012) 
assumed that the bound-free absorption can be neglected
even at late times
(i.e., the CSM is completely ionized).

Chandra et al.\ (2012a) reported on the analysis of
the X-ray spectrum of SN\,2010jl (including the proprietary data).
They measured a column density at their latest epoch of
$\sim10^{24}$\,cm$^{-2}$, which is about 1000
times larger than the Galactic column density in the direction
of SN\,2010jl.
Such a large bound-free absorption is expected
if the shock velocity is below $10^{4}$\,km\,s$^{-1}$
(i.e., the metals are not completely ionized; Chevalier \& Irwin 2012)
and the mass-loading parameter is as large as
expected from a wind shock breakout (e.g., Eq.~\ref{N}).

Moreover, the column density decreases by a factor of
$\sim3$ between  346 and 405 days after optical maximum light.
Equation~\ref{N} predicts that between these dates,
assuming $w=2$,
the bound-free absorption should decrease by only $\approx20$\%.
In order to explain such a fast decline in column density
over an $\approx20\%$ increase in time, we suggest that
in some regions above the shock breakout region
the CSM may have a steep density profile.
If indeed the CSM have
steep power-law index, we predict that the
free-free absorption will be low enough
and it may be possible to detect late-time radio
emission from this SN (see \S\ref{sec:radio}).
A complete analysis of all the available data,
including the proprietary data, is required
in order to understand the luminosity and spectral
evolution of this SN, and to give more firmer predictions.

\subsection{SN\,2006jc}
\label{SN2006jc}

The X-ray light curve of SN\,2006jc (Fig.~\ref{fig:SN2006jc_LCX}; Immler et al.\ 2008)
peaked around $\approx100$\,days after the explosion.
The visible-light rise time of SN\,2006jc
was shorter than about 15 days
(Foley et al.\ 2007; Pastorello et al.\ 2008).
Therefore, if this SN is powered by the diffusion
of shock energy 
and if we can approximate the shock breakout time scale as the
visible-light rise time,
then we can deduce that
the X-rays peaked $\gtorder 6t_{{\rm br}}$.
The emission line width in the spectra of SN\,2006jc
suggests that the shock velocity was likely below 4000\,km\,s$^{-1}$
(Foley et al.\ 2007).
If this estimate is correct, then it is possible that the
pre-shocked CSM of this SN is partially ionized
(Chevalier \& Irwin 2012).
This may explain the fact that the X-ray spectra
of SN\,2006jc become softer with time
(Fig.~\ref{fig:SN2006jc_LCX}; Immler et al.\ 2008).
As the shock propagates through the CSM
the column density and, therefore,
the bound-free absorption, both decreases.
This is consistent with the expectation that
the bound-free process will dominate
the absorption of soft X-rays,
and therefore, X-rays at 1\,keV will escape 
only after the column density drops below
$\approx4\times10^{21}$\,cm$^{-2}$.

Itagaki (2006) reported that a possible eruption,
with an absolute magnitude of about $-13$,
took place at the position of SN\,2006jc
about two years prior to the SN explosion
(see also Pastorello et al.\ 2008).
This pre-SN outburst of SN\,2006jc 
may have put in place a dense shell that could provide
the medium required for the formation of a collisionless shock.
Moreover, due to the eruptive nature of the event,
the outer edge of this shell may follows a density
profile that falls faster than $r^{-2}$, or has
a relatively sharp edge. Such a density profile
is required for the X-rays to escape
the shocked regions at times earlier
than the 10--50 shock-breakout time scale suggested
by Svirski et al.\ (2012).

\section{Conclusions}
\label{Conc}

We present a search for X-ray emission
from 28 SNe 
in which it is possible that 
the shock breakout took place within a dense CSM.
Most SNe have been observed with X-ray telescopes
only around maximum optical light.
The SNe in our sample that do have late time
observations, were either detected also
at early times or were observed serendipitously
at very late times.
In that respect,
our first conclusion is that a search
for X-ray emission both at early and late time
from SNe is essential to constrain the properties
of the CSM around their progenitors.

Our analysis suggest that some type-IIn/Ibn SNe,
most notably SN\,2006jc and SN\,2010jl,
have optical light curves
that are likely powered by a shock breakout
in CSM,
while some other type-IIn SNe do~not.
However, for most of the SNe in our sample,
including all the SLSN-I events,
the observations are not conclusive.
Specifically, the lack of X-ray detection
of SLSN-I events, cannot rule out
the interaction model suggested in Quimby et al.\ (2011a;
see also Ginzburg \& Balberg 2012).
We conclude that deeper observations at later times
are required in order to further test this model.

Given the limits found in this paper
and our current understanding of these events
it will be worthwhile to monitor
type-IIn SNe (as well as other classes of
potentially interacting SNe; e.g., type-IIL)
with X-ray and radio instruments at time scales
$\gtorder10$ times the rise time of the SN.
We argue that in some cases
bound-free absorption will play an important role
at early and late times.
Therefore, observations with the soon to be launched
Nuclear Spectroscopic Telescope Array (NuSTAR; Harrison et al.\ 2010)
in the 6--80\,keV band may be extremely useful to
test the theory and to study the physics of these
collisionless shocks.
Moreover, in the cases where bound-free absorption is important
(e.g., $v_{{\rm s}}\ltorder10^{4}$\,km\,s$^{-1}$; Chevalier \& Irwin 2012),
the spectral X-ray evolution as a function of time
can be use to probe the column density above the shock
at any given time,
and deduce the density profile outside the shocked regions.
We also argue that in some cases,
if the CSM has steep density profiles
(e.g., SN\,2010jl),
it may be possible to detect radio emission.

Finally, we note that Katz et al.\ (2011)
and Murase et al. (2011)
predict that the collisionless shocks
will generate TeV neutrinos.
These particles will be able to escape when
the collisionless shock begins.
The detection of such neutrinos
using Ice-Cube (Karle et al.\ 2003) will be a powerful tool to test
this theory and explore the physics of collisionless shocks.

\acknowledgments

We thank Ehud Nakar, Boaz Katz and Nir Sapir for many discussions.
This paper is based on observations obtained with the
Samuel Oschin Telescope as part of the Palomar Transient Factory
project, a scientific collaboration between the
California Institute of Technology,
Columbia University,
Las Cumbres Observatory,
the Lawrence Berkeley National Laboratory,
the National Energy Research Scientific Computing Center,
the University of Oxford, and the Weizmann Institute of Science.
EOO is incumbent of
the Arye Dissentshik career development chair and
is grateful to support by
a grant from the Israeli Ministry of Science.
MS acknowledges support from the Royal Society.
AC acknowledges support from LIGO, that was constructed by the
California Institute of Technology and Massachusetts Institute of
Technology with funding from the National Science
Foundation and operates under cooperative agreement PHY-0757058.
SRK and his group are partially supported by the
NSF grant AST-0507734.
AG acknowledges support by the Israeli, German-Israeli,
and the US-Israel Binational Science 
Foundations, 
a Minerva grant, and the Lord Sieff of Brimpton fund.
The National Energy Research Scientific Computing Center, which is supported by
the Office of Science of the U.S. Department of Energy under Contract No.
DE-AC02-05CH11231, provided staff, computational resources, and data storage for
the PTF project. PEN acknowledges support from the US Department of Energy Scientific
Discovery through Advanced Computing program under contract DE-FG02-06ER06-04.
JSB's work on PTF was supported by NSF/OIA award AST-0941742
(``Real-Time Classification of Massive Time-Series Data Streams").
LB is supported by the NSF under
grants PHY 05-51164 and AST 07-07633.
MMK acknowledges generous support from the Hubble Fellowship and
Carnegie-Princeton Fellowship.

\end{document}